\documentclass[journal]{IEEEtran}

\usepackage{xcolor}
\usepackage{graphicx}
\usepackage{tikz}
\usepackage{color}
\usepackage{amsmath}
\usepackage{subfig}
\usepackage{algorithm}
\usepackage{algorithmic}
\usepackage{amssymb}
\usepackage{math}
\usepackage{caption}
\usepackage{array}

\begin{document}
\bibliographystyle{ieeetr}

\title{Multichannel Speech Separation and Enhancement  Using the  Convolutive Transfer Function}
\author{Xiaofei Li, Laurent Girin, Sharon Gannot and Radu Horaud
\thanks{X. Li and R. Horaud are with INRIA Grenoble Rh\^one-Alpes, Montbonnot Saint-Martin, France. }
\thanks{L. Girin is with GIPSA-lab and with Univ. Grenoble Alpes, Saint-Martin d'H\`eres, France.}
\thanks{Sharon Gannot is with Bar Ilan University, Faculty of Engineering, Israel. }
\thanks{This work was supported by the ERC Advanced Grant VHIA \#340113.}
}

\maketitle

\begin{abstract}
This paper addresses the problem of speech separation and enhancement from multichannel convolutive  and noisy mixtures, \emph{assuming known mixing filters}. We propose to perform the speech separation and enhancement task in the short-time Fourier transform domain, using the convolutive transfer function (CTF) approximation. 
Compared to time-domain filters, CTF has much less taps, consequently it has less near-common zeros among channels and less computational complexity.
The work proposes three speech-source recovery methods, namely: i) the multichannel inverse filtering method, i.e. the multiple input/output inverse theorem (MINT), is exploited in the CTF domain, and for the multi-source case,  
ii) a beamforming-like multichannel inverse filtering method applying single source MINT and using power minimization, which is suitable whenever the source CTFs are not all known, and
iii) a constrained Lasso method, where the sources are recovered by minimizing the $\ell_1$-norm to impose  their spectral sparsity, with the constraint that the $\ell_2$-norm fitting cost, between the microphone signals and the mixing model involving the unknown source signals, is less than a tolerance. The noise can be reduced by setting a tolerance onto the noise power.
Experiments under various acoustic conditions are carried out to evaluate the three proposed methods. The comparison between them as well as with the baseline methods is presented. 
\end{abstract}


\section{Introduction}
\label{sec:introduction}
Speech recordings in the real world consist of the convolutive images of multiple audio sources and some additive noise. 
A convolutive image is the convolution between the source signal and the room impulse response (RIR), which is also called mixing filter in the multisource context. Correspondingly, the distortions on the source signals, i.e. interfering speakers, reverberations and additive noise, heavily deteriorate the speech intelligibility for both human listening and machine recognition.
This work aims to suppress these distortions, in other words, to recover the respective source signals from the multichannel recordings. 
In general,  suppressing interfering speakers, reverberations and noise are respectively refered to source separation, dereverberation and noise reduction. 
Each of which is a difficult task, that attracts lots of research attentions. 
In the microphone recordings, there are three unknown terms, i.e. source signals, mixing filters, and noise. 
Thence, the problem is often split into two subproblems i) identification of mixing filters and noise statistics, and ii) estimation of the source signals.
This work focuses on the problem of speech source estimation assuming that the mixing filters, and possibly the noise statistics, are either known or their estimates are available.

Most convolutive source separation and speech enhancement techniques are designed in the short time Fourier transform (STFT) domain. 
In this domain, the convolutive process is usually  approximated at each time-frequency (TF) bin by a product between the source STFT coefficient and the Fourier transform of the mixing filter. 
This assumption is called the multiplicative transfer function (MTF) approximation \cite{avargel2007spl}, or the narrowband approximation, and the frequency domain mixing filter is called the acoustic transfer function (ATF).
Based on the known ATFs, or the respective relative transfer functions (RTFs) \cite{gannot2001,li2015icassp}, the beamforming techniques are widely used for multichannel source separation and speech enhancement, such as the minimum variance/power distortionless response (MVDR/MPDR) beamformer, and the linearly constrained minimum variance/power (LCMV/LCMP) beamformer \cite{gannot2001,van2004}.  
Moreover, the sparsity of the audio signals in the TF domain can be utilized. 
Based on this property, the binary masking \cite{yilmaz2004,mandel2010} and the $\ell_1$-norm minimization \cite{winter2007} approaches have been applied for source separation.
For more examples of MTF-based techniques, please refer to a comprehensive review \cite{gannot2017} and references therein.

The narrowband assumption is theoretically valid only if the length of the mixing filters is small relative to the length of the STFT window. 
In practice, this is very rarely the case, even for moderately reverberant environments, since the STFT window is limited to assume local stationarity of audio signals.
Hence the narrowband assumption fundamentally hamper the speech enhancement performance, and this becomes critical for strongly reverberant environments. 
To avoid the limitation of narrowband assumption, several source separation methods based on the time-domain representation of mixing filters have been proposed. 
In the wide-band Lasso method \cite{kowalski2010}, the source signals are estimated by minimizing an  $\ell_2$-norm fitting cost between the microphone signals and the mixing model involving the unknown source signals, in which the exact time-domain (wide-band) source-filter convolution is used. 
Importantly, the $\ell_1$-norm of the STFT-domain source signals is  added to the fitting cost as a regularization term to impose the spectral sparsity of the source spectra. 
In the presence of additive noise, the $\ell_1$-norm regularization is able to reduce the noise in the recovered source signals. 
However, the regularization factor is difficult to set even if the noise power is known. To overcome this, a more flexible scheme is proposed in \cite{arberet2013} that relaxes the $\ell_2$-norm fitting cost to the noise level and minimizes the $\ell_1$-norm. 
In addition, a reweighting approach is also proposed in \cite{arberet2013} to approximate the $\ell_0$-norm. In the family of multichannel inverse filtering or multichannel equalization, an inverse filter is estimated with respect to the known mixing filters, and applied to the microphone signals, preserving the desired source and suppressing the interfering sources.
The multiple-input/output inverse theorem (MINT) method \cite{miyoshi1988} was first proposed for this aim, which however is sensitive to RIR perturbations (misalignment / estimation error) and to microphone noise. 
To improve the robustness of MINT to RIR perturbations, many techniques have been proposed, preserving not only the direct-path impulse response but also the early reflections, such as channel shortening \cite{kallinger2006}, infinity- and $p$-norm optimization-based channel shortening/reshaping \cite{mertins2010}, partial MINT \cite{kodrasi2013,kodrasi2016}, etc. In addition, the energy of the inverse filter was used in \cite{hikichi2007} as a regularization term to avoid the amplification of filter perturbations and microphone noise. In \cite{huang2005}, a two-stage method was proposed, that first converts a multiple-input multiple-output (MIMO) system to multiple single-input multiple-output (SIMO) systems for source separation, and then applies inverse filtering for dereverberation.

The wide-band models mentioned above are all performed in the time domain.
The time-domain convolution problem can be transformed to the subband domain, which provides several benefits i) the original problem is split into subproblems, and each subproblem has a smaller data size and thus a smaller computational complexity, ii) the subband mixing filters are shorter than the time-domain filters, thence are likely to have less near-common zeros among microphones, which benefits both the filter identification and the multichannel equalization, even if the former is beyond the scope of this work, and    
iii) in the TF domain, the sparsity of the speech signal can be more easily exploited.  Several variants of subband MINT were proposed based on filter banks \cite{yamada1991,wang1992,weiss1999,gaubitch2009,lim2013}. The key issues in the filter-bank design are i) the time-domain RIRs should be well approximated in the subband  domain, and ii) the frequency response of each filter-bank should be fully excited, i.e. should not involve the frequency components with the magnitude close to zero. 
Otherwise, these components are common to all channels, and are problematic in  the MINT application. To satisfy the second condition, the filter-bank is either critically sampled \cite{yamada1991,wang1992}, which suffers from frequency aliasing, or has a flat-top frequency response  \cite{weiss1999,gaubitch2009,lim2013}, which may suffer from  time aliasing.
Generally speaking, the STFT transform is more preferable in the sense that most of the acoustic algorithms in the current literature are performed in this domain. To represent the time-domain convolution in the STFT domain, especially for the long filter case,  cross-band filters were introduced in \cite{avargel2007}. To simplify the analysis, the convolutive transfer function (CTF) approximation is further adopted in \cite{talmon2009,talmon2009convolutive} only using the band-to-band convolution and ignoring the cross-band filters. 
In \cite{talmon2009convolutive}, CTF is integrated into the generalized sidelobe canceler beamformer. In our previous works \cite{li2016taslp} and \cite{li2017taslp}, blindly estimated CTF, specifically its direct-path part, was used for localizing single speaker and multiple speakers, respectively. In \cite{li2017icassp}, a CTF-Lasso method was proposed following the spirit of the wide-band Lasso \cite{kowalski2010}.

Several probabilistic techniques have also been proposed for wide-band source separation via maximizing the likelihood of a generative model. 
Variational Expectation-Maximization (EM) algorithms are proposed in \cite{leglaive2017} and \cite{leglaive2017separating} based on the time-domain convolution and in \cite{badeau2014} based on cross-band filters. 
CTF-based EM algorithms are proposed in \cite{schwartz2015online} and \cite{li2017waspaa} for single source dereverberation and source separation, respectively. These EM algorithms iteratively estimate the mixing filters and the sources, and intrinsically require a fairly good initialization for both filters and sources. 

In this work, we propose the following three source recovery methods in the standard oversampled STFT domain using the CTF approximation: 
\begin{itemize}
 \item All the above-mentioned improved MINT methods are proposed for single source dereverberation. The multisource case has been rarely studied, even if the multisource MINT was presented in the original paper \cite{miyoshi1988}. We propose a CTF-based multisource MINT method for both source separation and dereverberation. The oversampled STFT does not suffer from both frequency aliasing and time aliasing. However, the STFT window is not flat-top, namely the subband signals and filters have a frequency region with a magnitude close to zero, which is common to all channels. To overcome this problem, instead of using the conventional impulse function as the target of the inverse filtering, we propose a new target, which has a frequency response corresponding to the STFT window.
In addition,  a filter energy regularization is adopted following \cite{hikichi2007} to improve the robustness of inverse filtering. 
\item For situations where the CTFs of the sources are not all available, we propose a beamforming-like inverse filtering method. The inverse filters are designed i) to preserve one source with known CTFs based on single source MINT, and ii) to minimize the overall power of the inverse filtering output, and thus suppress the interfering sources and noise. This method shares a similar spirit with the MPDR beamformer. 
\item To overcome the drawback of the CTF-Lasso method \cite{li2017icassp}, namely that the regularization factor is difficult to set with respect to the noise level, following the spirit of \cite{arberet2013}, 
we propose to recover the source signals by minimizing the $\ell_1$-norm of the source spectra with the constraint that the $\ell_2$-norm fitting cost is less than a tolerance. 
The setting of the tolerance is studied. In addition, a complex-valued \emph{proximal splitting} algorithm \cite{combettes2007,combettes2005} is investigated to solve the optimization problem. 
 \end{itemize}

The remainder of this paper is organized as follows. The problem is formulated based on CTF in Section~\ref{sec:formulation}. 
The two multichannel inverse filtering methods are proposed in Section~\ref{sec:if}.
The improved CTF-Lasso method is proposed in  Section~\ref{sec:lasso}. Experiments are presented in Section~\ref{sec:experiments}.  
Section~\ref{sec:conclusion} concludes the work.

\section{CTF-based Problem Formulation} \label{sec:formulation}

In the time domain, we consider a multichannel convolutive mixture with $J$ sources and $I$ microphones,
\begin{align}\label{eq:xn}
 x^i(n)  = \sum_{j=1}^J a^{i,j}(n) \star s^j(n)+e^i(n),  
 \end{align}
where $n$ is the time index, and $i=1,\dots,I, \ I\ge2$ and $j=1,\dots,J, \ J\ge2$ are respectively the indices of the microphones and the sources. 
The signals $x^i(n)$, $s^j(n)$ and $e^i(n)$ are microphone signals, source signals, and noise signals, respectively. Here $\star$ denotes convolution, and $a^{i,j}(n)$ is the RIR relating the $j$-th source to the $i$-th microphone.  
Note that the relation between $I$ and $J$ is not specified here, and this will be discussed afterwards with respect to the proposed methods. 
The noise signals $e^i(n)$ are uncorrelated with the source signals, and could be spatially uncorrelated, diffuse, or directional.  
 
The goal of this paper is to recover the multiple source signals from the microphone signals, given the RIRs and the noise PSDs. 
The RIRs and noise PSDs could be blindly estimated from the microphone signals, and the estimated values generally suffer from disturbances, which are not trivial but beyond the scope of this work. 
Overall, the multi-source recovery problem implies that  source separation, dereverberation, and noise reduction are  conducted  simultaneously. 


\subsection{Convolutive Transfer Function}
In this section, the time-domain convolution is transformed into the STFT-domain CTF convolution. 
To simplify the exposition, we consider, for the meantime, the noise free situation with only one microphone and one source: $x(n) = a(n)\star s(n)$, where the source and microphone indices are omitted. 

The STFT representation of the microphone signal $x(n)$ is 
\begin{align}\label{eq:stft}
 x_{p,k} = \sum_{n=-\infty}^{+\infty} x(n)\tilde{w}(n-pD)e^{-j\frac{2\pi}{N}k(n-pD)},
\end{align}
where $p$ and $k$ denote the frame index and the frequency index, respectively. $\tilde{w}(n)$ is the STFT analysis window, and $N$ and  $D$ denote the frame (window) length, and the frame step, respectively.
In the filter bank interpretation, the analysis window is considered as the low-pass filter, and $D$ as the decimation factor.

The cross-band filter model \cite{avargel2007} consists in representing the STFT coefficient $x_{p,k}$ as a summation over multiple convolutions (between the STFT-domain source signal $s_{p,k}$ and filter $a_{p,k,k'}$) across frequency bins.
Mathematically, the linear time invariant system can be written in the STFT domain as 
\begin{align}\label{xpk2}
 x_{p,k} = \sum_{k'=0}^{N-1}\sum_{p'} s_{p-p',k'} \; a_{p',k,k'},
\end{align}
If $D<N$, then $a_{p',k,k'}$ is non-causal, with $\lceil N/D \rceil -1$ non-causal coefficients, where $\lceil \cdot \rceil$ denotes the ceiling function. The number of causal filter coefficients is related to the reverberation time. For notational simplicity, let the filter index $p'$ be in $[0,L_a-1]$, with $L_a$ being the filter length, i.e. the non-causal coefficients are shifted to the causal part, which only leads to a constant shift of the frame index of the source signal.
Let $w(n)$ denote the STFT synthesis window.
The STFT-domain impulse response $a_{p',k,k'}$ is related to the time-domain impulse response $a(n)$ by:
\begin{align}\label{eq:hp}
a_{p',k,k'}={(a(n)\star \zeta_{k,k'}(n))}|_{n=p'D},
\end{align}
which represents the convolution with respect to the time index $n$ evaluated at frame steps, with
\begin{align}
\zeta_{k,k'}(n) = e^{j\frac{2\pi}{N}k'n}\sum_{m=-\infty}^{+\infty} \tilde{w}(m) \: w(n+m) \: e^{-j\frac{2\pi}{N}m(k-k')}. \nonumber
\end{align}
To simplify the analysis, we consider the CTF approximation, i.e., only band-to-band filters with $k=k'$ are considered: 
\begin{equation}
\label{eq:xpk3}
 x_{p,k} \approx \sum\nolimits_{p'=0}^{L_a-1} s_{p-p',k}a_{p',k}= s_{p,k}\star a_{p,k}.
 \end{equation}

\subsection{STFT Domain Mixing Model} 

Based on the CTF approximation, we can obtain the STFT-domain mixing model corresponding to the time-domain model~(\ref{eq:xn}), 
\begin{align}\label{eq:xpi}
 x_{p}^i = \sum_{j=1}^J a_{p}^{i,j} \star s_{p}^j+e_{p}^i,  
 \end{align}
Note that here (and hereafter) the frequency index $k$ is omitted, unless it is necessary. Since the proposed methods are applied frequency-wise. Let $p\in[1,P]$ and $p\in[0,L_a-1]$  denote the frame indices of the microphone signals and the CTFs respectively. The goal of this work is to recover the STFT coefficients of the source signals, i.e. $s_{p}^j$, and then applying the inverse STFT to obtain an estimation of the time-domain source signals.

\section{multichannel Inverse Filtering} \label{sec:if}

The multichannel inverse filtering method is based on the MINT method. In this section, we propose two MINT-based methods in the CTF domain for the multisource case.

\subsection{Problem Formulation for Inverse Filtering}
Define the CTF-domain inverse filters as $h_p^{i}$ with $i=1,\dots,I$ and $p=0,\dots,L_h-1$, where $L_h$ denotes the length of the inverse filters. The output of the inverse filtering is 
\begin{align}\label{eq:out}
 y_p  = \sum_{i=1}^{I} h_p^{i} \star x_p^i  =\sum_{j=1}^J s_p^j \star \left(\sum_{i=1}^{I} h_p^{i} \star a_p^{i,j}\right) + \sum_{i=1}^{I} h_p^{i} \star e_p^i, 
\end{align}
which comprises the mixture of the inverse filtered sources  and the inverse filtered noise.

To facilitate the analysis, we denote the convolution in vector form. We define the convolution matrix for the microphone signal $x_p^i$ as:
\begin{equation}\label{eq:cm}
\mathbf{X}^{i} =
\begin{bmatrix}
x_1^i & 0     &  \cdots   & 0  \\
x_2^i & x_1^i &  \ddots  & \vdots     \\
\vdots & \ddots &   \ddots & \vdots     \\ 
x_{P}^i & \vdots & \ddots & 0   \\ 
0 & x_{P}^i & \ddots & \vdots \\
\vdots & \ddots & \ddots    & \vdots \\
0& \cdots &0  & x_{P}^i \\
\end{bmatrix} \in \mathbb{C}^{(P+L_h-1)\times L_h},  
\end{equation}
and the vector of filter $h_p^i$ as 
\begin{align}
 \mathbf{h}^i = [{h}_0^{i},\dots,{h}_p^{i},\dots,{h}_{L_h-1}^{i}]^{\top} \in \mathbb{C}^{L_h\times 1}, \nonumber
\end{align}
where $^{\top}$ denotes the vector or matrix transpose. Then the convolution 
$h_p^{i} \star x_p^i$ can be written as $\mathbf{X}^{i}\mathbf{h}^i$.
The inverse filtering~(\ref{eq:out}) can be written as:
\begin{align}\label{eq:outvect}
 \mathbf{y} = \mathbf{X}\mathbf{h},
\end{align}
with: 
\begin{align}
\mathbf{y} & = [y_1,\dots,y_p,\dots,y_{P+L_h-1}]^{\top} \in \mathbb{C}^{(P+L_h-1)\times 1},  \nonumber \\
\mathbf{X} & = [\mathbf{X}^{1},\dots,\mathbf{X}^{i},\dots,\mathbf{X}^{I}] \in \mathbb{C}^{(P+L_h-1)\times IL_h},  \nonumber  \\
\mathbf{h} & = [\mathbf{h}^{1{\top}},\dots,\mathbf{h}^{i{\top}},\dots,\mathbf{h}^{I{\top}}]^{\top} \in \mathbb{C}^{IL_h\times 1}. \nonumber
\end{align}

Similarly, we define the convolution matrix for the CTF $a_p^{i,j}$ as $\mathbf{A}^{i,j}\in \mathbb{C}^{(L_a+L_h-1)\times L_h}$, and write $h_p^{i} \star a_p^{i,j}$ as $\mathbf{A}^{i,j}\mathbf{h}^i$. Moreover, we define $\mathbf{A}^j=[A^{1,j},\dots,A^{i,j},\dots,A^{I,j}]\in\mathbb{C}^{(L_a+L_h-1)\times IL_h}$, and write $\sum_{i=1}^{I} h_p^{i} \star a_p^{i,j}$ as $\mathbf{A}^j \mathbf{h}$.

\subsection{The CTF-MINT Formulation}

To preserve a desired source, e.g. the $j_d$-th source, the inverse filtering of the CTF filters, i.e. $\sum_{i=1}^{I} h_p^{i} \star a_p^{i,j_d}$, should target an impulse function function ${d}_p$ with  length  $L_a+L_h-1$. To suppress the interfering sources, the inverse filtering of the CTF filters of the other sources, i.e. $\sum_{i=1}^{I} h_p^{i} \star a_p^{i,j\ne j_d}$, should target a zero signal. 
Let $\mathbf{d}$ denote the vector form of $d_p$, and $\mathbf{0}$ denote a ($L_a+L_h-1$)-dimensional zero vector.
We define the following $I$-input $J$-output MINT equation
\begin{equation}
\begin{bmatrix}
 \mathbf{0} \\
 \vdots \\
 \mathbf{0} \\
 \mathbf{d} \\
  \mathbf{0} \\
 \vdots \\
 \mathbf{0} 
\end{bmatrix}
=
\begin{bmatrix}
\mathbf{A}^{1,1} &  \cdots & \mathbf{A}^{I,1}   \\
\vdots & \ddots &\vdots  \\
\mathbf{A}^{1,j_d-1} & \cdots &\mathbf{A}^{I,j_d-1}    \\ 
\mathbf{A}^{1,j_d} &\cdots & \mathbf{A}^{I,j_d}    \\ 
\mathbf{A}^{1,j_d+1}&\cdots & \mathbf{A}^{I,j_d+1} \\
\vdots & \ddots &\vdots  \\
\mathbf{A}^{1,J} &\cdots &\mathbf{A}^{I,J}  
\end{bmatrix}
\begin{bmatrix}
 \mathbf{h}^1 \\
 \vdots \\
 \mathbf{h}^I 
\end{bmatrix}
=
\begin{bmatrix}
\mathbf{A}^1  \\
\vdots    \\
\mathbf{A}^{j_d-1}     \\ 
\mathbf{A}^{j_d}   \\ 
\mathbf{A}^{j_d+1} \\
\vdots \\
\mathbf{A}^J 
\end{bmatrix} 
\mathbf{h} \nonumber
\end{equation}
which can be rewritten in a compact form as
\begin{align}\label{eq:mint}
 \mathbf{g}= \mathbf{A}\mathbf{h}. 
\end{align}
When the matrix $\mathbf{A}\in\mathbb{C}^{J(L_a+L_h-1)\times IL_h}$ is either square or wide, namely $IL_h \ge J(L_a+L_h-1)$ and thus $L_h\ge\frac{J(L_a-1)}{I-J}$, (\ref{eq:mint}) has an exact solution, which means an exact inverse filtering can be achieved.   
This condition implies an over-determined recording system, i.e. $I>J$. 

From \cite{miyoshi1988}, the solvable condition of (\ref{eq:mint}) is that the CTFs of the desired source $a_p^{i,j_d}, i=1,\dots,I$, do not have any common zero. On one hand, the subband filters, i.e. the CTFs, are much shorter than the time-domain filters, and are thus likely to have much less near-common zeros, which is a major benefit. On the other hand, the filter banks induced from the short-time windows lead to some structured common zeros.
From (\ref{eq:hp}), for any RIR $a^{i,j}(n)$, its CTF (with $k'=k$) is computed as
\begin{align}\label{eq:apk}
a_{p,k}^{i,j}={(a^{i,j}(n)\star \zeta_{k}(n))}|_{n=pD},
\end{align}
with 
\begin{align}
\zeta_{k}(n) = e^{j\frac{2\pi}{N}kn}\sum_{m=-\infty}^{+\infty} \tilde{w}(m) \: w(n+m) \nonumber
\end{align}
being the cross-correlation of the analysis window $\tilde{w}(n)$ and the synthesis window $w(n)$ modulated (frequency shifted) by $e^{j\frac{2\pi}{N}kn}$. 
This cross-correlation has a similar frequency response as the windows $\tilde{w}(n)$ and $w(n)$ in the sense that it is also a low-pass filter with the same bandwidth denoted by $\bar{\omega}$.
The frequency response of $a_{p,k}^{i,j}$ is the frequency response of $a^{i,j}(n)$ multiplied by the frequency response of $\zeta_{k}(n)$, and then folded by downsampling with a period of $2\pi/D$. 
To avoid frequency aliasing, the period should not be smaller than the bandwidth $\bar{\omega}$ not to fold the passband of the low-pass filter. For example, in this work, we use the Hamming window, the width of the main lobe is considered as the bandwidth, i.e. $\bar{\omega}=8\pi/N$. Consequently, we set the constraint $D\le N/4$. If we consider the magnitude of side lobes to be zero, the frequency response of $a_{p,k}^{i,j}$ can be interpreted as the $k$-th frequency band of $a^{i,j}(n)$ multiplied by the frequency response of the downsampled $\zeta_{k}(n)$, i.e. $\zeta_{p,k}=\zeta_{k}(n)|_{n=pD}$.
When $D< N/4$, the frequency response of $\zeta_{p,k}$ involves some side lobes, which have a magnitude close to zero. When $D=N/4$, only the main lobe is involved, and because the magnitude is dramatically decreasing from the center of the main lobe to its margin, the frequency region close to the margin of the main lobe has  magnitude close to zero. 
This phenomenon, namely that the frequency response of $\zeta_{p,k}$ and thus of $a_{p,k}^{i,j}$ are not fully excited, is common to all microphones, which is problematic for solving (\ref{eq:mint}).
Fortunately, it is trivially known that the common zeros are introduced by the frequency response of $\zeta_{p,k}$. 
To make (\ref{eq:mint}) solvable, we propose to determine the desired target $\mathbf{d}$ to have the same frequency response as $\zeta_{p,k}$, instead of the impulse function that has a full-band frequency response. 
To this end, the target $\mathbf{d}$ is designed as:
\begin{align}\label{eq:d}
 \mathbf{d} = [0,\dots,0,\zetavect^{\top},0,\dots,0]^{\top} \in\mathbb{C}^{(L_a+L_h-1)\times 1},
\end{align}
where $\zetavect$ denotes the vector form of $\zeta_{p,k}$. 
The zeros before $\zetavect$ introduce a modeling delay. As shown in \cite{hikichi2007}, this delay is important for making the inverse filtering robust to perturbations of the CTF.     

The solution of (\ref{eq:mint}) gives an exact recovery of the $j_d$-th source plus the filtered noise $\sum_{i=1}^{I} h_p^{i} \star e_p^i$ as shown in (\ref{eq:out}). In this method, a directional noise can be treated as an interfering source, and be modeled in the MINT formulation. 
Therefore, here we only need to consider the spatially uncorrelated or diffuse noise $e_p^i$.
To suppress the noise, a straightforward way is to minimize the power of the filtered noise under the MINT constraint (\ref{eq:mint}).
As proposed in \cite{hikichi2007}, an alternative way to suppress the noise is to reduce the energy of the inverse filter $\mathbf{h}$. 
This strategy is equivalent to minimizing the power of the filtered noise if we approximately assume the noise correlation matrix is the identity. 
In addition, this strategy is also capable to suppress the perturbations of the CTFs, if the disturbance noise is also assumed to have an identity correlation matrix. This leads to the following optimization problem: 
\begin{align}\label{eq:ctf-mint}
 \mathop{\textrm{min}}_{\mathbf{h}}  \parallel \mathbf{A}\mathbf{h}-\mathbf{g} \parallel^2 + \delta \phi_a^{j_d} \parallel \mathbf{h} \parallel^2,  
\end{align}
where $\phi_a^{j_d}=\sum_{i=1}^{I}\sum_{p=0}^{L_a-1}|a_p^{i,j_d}|^2$ is the CTF energy for the desired source (summed over channels and frames), used as a normalization term, and $\delta$ is the regularization factor.
Indeed, the power of the inverse filter $\mathbf{h}$ is at the level of $1/\phi_a^{j_d}$, thus $\parallel \! \mathbf{h} \! \parallel^2$ is somehow normalized by $\phi_a^{j_d}$. 
As a result, the choice of $\delta$, which controls the trade-off between the two terms in (\ref{eq:ctf-mint}), is made independent of the energy level of the CTF filters. 
This property is especially relevant for the present frequency-wise algorithm since all frequencies can share the same regularization factor $\delta$, although the CTF energy may significantly vary along the frequencies.
The solution of (\ref{eq:ctf-mint}), i.e. the  CTF-based regularized MINT inverse filter, is
\begin{align}\label{eq:sctf-mint}
 \hat{\mathbf{h}}^{\text{mint}} = (\mathbf{A}^H\mathbf{A}+\delta \phi_a^{j_d} \mathbf{I})^{-1}\mathbf{A}^H\mathbf{g},
\end{align}
where $\mathbf{I}$ is the $IL_h$-dimensional identity matrix.
We refer to this method as CTF-MINT.

As mentioned above, to perform the exact inverse filtering, matrix $\mathbf{A}$ should be either square or wide. 
In (\ref{eq:ctf-mint}), the exact match between $\mathbf{A}\mathbf{h}$ and $\mathbf{g}$ is relaxed, which means the exact inverse filtering is abandoned to improve the robustness of the inverse filter estimate.  
Let $\rho$ denote the ratio between the number of columns and the number of rows of $\mathbf{A}$, then we have $IL_h = \rho J(L_a+L_h-1)$. Rename $L_h$ as $L_h^{\text{mint}}$, then:   
\begin{align}\label{eq:lhmint}
 L_h^{\text{mint}}=\frac{L_a-1}{\frac{I}{\rho J}-1}, \quad \text{ with } \rho<\frac{I}{J}.
\end{align}
For the over-determined recording system, i.e. $I>J$, we can set $\rho\ge1$ to have a square or wide $\mathbf{A}$. When $I\le J$, $\rho$ should be less than $\frac{I}{J}$, consequently $\mathbf{A}$ is narrow,  
however, as opposed to solving (\ref{eq:mint}), the optimization problem (\ref{eq:ctf-mint}) is still feasible. 
Note that $L_h^{\text{mint}}\rightarrow +\infty$ when $\rho\rightarrow \frac{I}{J}$, thence in practice $\rho$ should be sufficiently small to avoid a very large $L_h^{\text{mint}}$.

\subsection{The CTF-MPDR Formulation}

The above CTF-MINT approach requires CTF knowledge of all the sources. In this section, we consider the situation where the CTFs of the sources are not all obtained/estimated. One source is recovered based on  its own CTFs  only.

For the desired source, the inverse filter $\mathbf{h}$ should still satisfy $\mathbf{A}^{j_d}\mathbf{h}=\mathbf{d}$ to achieve a distortionless desired source. 
At the same time, the power of the output, i.e. $\parallel \mathbf{X}\mathbf{h} \parallel^2$, should be minimized. Again, by relaxing the match between $\mathbf{A}^{j_d}\mathbf{h}$ and $\mathbf{d}$, we define the following optimization problem
\begin{align}\label{eq:ctf-mpdr}
 \mathop{\textrm{min}}_{\mathbf{h}}  \parallel \mathbf{A}^{j_d}\mathbf{h}-\mathbf{d} \parallel^2 + \kappa \frac{\phi_a^{j_d}}{\phi_x} \parallel \mathbf{X}\mathbf{h} \parallel^2,  
\end{align}
where $\phi_x = \sum_{i=1}^{I}\sum_{p=0}^{P-1}|x_p^{i}|^2$ is the energy of the microphone signals. 
Similar to CTF-MINT, the normalization factor $\frac{\phi_a^{j_d}}{\phi_x}$ makes the choice of the regularization factor $\kappa$ independent of the energy of the CTF filters and the energy of the microphone signals. 
Therefore, all the frequencies can share the same regularization factor $\kappa$, even if the energy of microphone signals significantly varies across frequencies.  This optimization problem considers any type of noise signal equally by minimizing the overall output power. 

The solution of (\ref{eq:ctf-mpdr}), i.e. the  CTF-based beamforming-like inverse filter, is
\begin{align}\label{eq:sctf-mpdr}
 \hat{\mathbf{h}}^{\text{mpdr}} = (\mathbf{A}^{j_dH}\mathbf{A}^{j_d}+\kappa \frac{\phi_a^{j_d}}{\phi_x} \mathbf{X}^H\mathbf{X})^{-1}\mathbf{A}^{j_dH}\mathbf{d}.
\end{align}
This method is similar in spirit with the  MPDR beamformer, 
more exactly with the speech distortion weighted multichannel Wiener filter \cite{doclo2005} since the source distortionless is relaxed. We still refer to this method as CTF-MPDR.

Similarly, let $\varrho$ denote the ratio between the number of columns and the number of rows of $\mathbf{A}^{j_d}$, then we have $IL_h = \varrho(L_a+L_h-1)$. Rename $L_h$ as $L_h^{\text{mpdr}}$, then 
\begin{align}\label{eq:lhmpdr}
 L_h^{\text{mpdr}}=\frac{L_a-1}{\frac{I}{\varrho}-1}, \quad \text{ with } \varrho<I.
\end{align}
Because the inverse filter is constrained by only one source, i.e. the desired source, it can always be set as $\varrho\ge 1$ in order to have either square or wide $\mathbf{A}^{j_d}$.

For both CTF-MINT and CTF-MPDR, the $J$ source signals are estimated by respectively taking the $1,\cdots,J$-th source as the desired source and appling (\ref{eq:out}). They both do not require the knowledge of noise statistic.

\section{CTF-based Constrained Lasso}\label{sec:lasso}

Instead of explicitly estimating an inverse filter, the source signals can be directly recovered by matching the microphone signals and the mixing model involving the unknown source signals.
To this end, the spectral sparsity of the speech signals could be exploited as \emph{prior} knowledge. 

\subsection{Problem Formulation for the Mixing model}
The mixing model (\ref{eq:xpi}) can be rewritten in vector/matrix form as
\begin{align}\label{eq:bfx}
 \mathbf{x}=\mathcal{A}\star\mathbf{s}+\mathbf{e},
\end{align}
where $\mathbf{x}\in \mathbb{C}^{I\times P}$, $\mathbf{s}\in \mathbb{C}^{J\times P}$ and $\mathbf{e}\in \mathbb{C}^{I\times P}$ denote the matrices of microphone signals, source signals and noise signals, respectively, 
and $\mathcal{A}\in \mathbb{C}^{I\times J\times P}$ denotes the three-way CTF array. The convolution $\star$ is carried out along the time frame. Remember that this equation is defined for each frequency bin $k$ and that we omit the $k$ index for clarity of presentation.
In Section \ref{sec:if}, the convolution between two signals was formulated as the multiplication of the convolution matrix of one signal and the vector form of the other signal. 
In the present section, the convolution operator $\star$ is considered in its conventional form. The reason is that, in the  method proposed here, only the convolution operation itself is used, which can be achieved by the fast Fourier transform.

In our previous work \cite{li2017icassp}, we proposed to estimate the source signals by solving an $\ell_2$-norm fitting cost minimization problem with an $\ell_1$-norm regularization term
\begin{align}\label{eq:lasso}
\mathop{\textrm{min}}_{\mathbf{s}} \parallel \mathcal{A}\star\mathbf{s} - \mathbf{x} \parallel^2+\lambda | \mathbf{s} |,
\end{align}
where $\lambda$ is the regularization factor. Note that both the $\ell_2-$ and $\ell_1$-norms on matrices are redefined here as vector norms. 
The first term minimizes the fitting cost, and the second term imposes sparsity on the speech source signals.
In the presence of additional noise $\mathbf{e}$, the regularization factor $\lambda$ can be adjusted to impose the sparsity and thus to remove the noise from the estimated source signals.
However, it is difficult to automatically tune $\lambda$ even when the noise PSD is known. 
Especially, the source  recovery is performed frequency by frequency in this work, and it is common that the noise PSD
has different values at different frequencies. 
This requires a specific value of $\lambda$ for each frequency, which further increases the difficulty of choosing $\lambda$. In this work, we solve this problem by transforming the above problem to a constrained optimization problem. 

\subsection{CTF-based Constrained Lasso}

Problem (\ref{eq:lasso}) is equivalent to the following formulation
\begin{align}\label{eq:classo}
\mathop{\textrm{min}}_{\mathbf{s}}   | \mathbf{s} |, \quad \text{s.t. }   \parallel \mathcal{A}\star\mathbf{s} - \mathbf{x} \parallel^2 \le \epsilon, 
\end{align}
for some unknown $\lambda$ and $\epsilon$. The $\ell_2$-norm fitting cost is relaxed to at most a tolerance $\epsilon$. This formulation was first proposed in \cite{arberet2013} for audio source separation in the time domain. 
We adapted it to the CTF-magnitude domain in our previous work \cite{li2017blind} for single source dereverberation. In the present work, we further extend it to the complex-valued CTF domain for multisource recovery.

The setting of the tolerance $\epsilon$ is critical to the quality of the recovered source signals.
The tolerance $\epsilon$ is related to the noise power in the microphone signals. The noise signal is assumed to be stationary. 
Let $\sigma_{i}^2$ denote the noise PSD in the $i$-th microphone, which can be estimated from pure noise signal or estimated by a noise PSD estimator,  e.g. \cite{li2016icassp}.  
Let $\mathbf{e}^i\in \mathbb{C}^{1\times P}$ denote the noise signal in the $i$-th microphone in vector form.  
The squared $\ell_2$-norm of the noise signal, i.e. the noise energy $\parallel\mathbf{e}^i\parallel^2$, follows an Erlang distribution with mean $P\sigma_{i}^2$ and variance $P\sigma_{i}^4$ \cite{forbes2010}. 
We assume that noise signals are spatially uncorrelated, then for all microphones, the squared $\ell_2$-norm $\parallel\mathbf{e}\parallel^2$ has mean $\sum_{i=1}^{I}P\sigma_{i}^2$ and variance $\sum_{i=1}^{I}P\sigma_{i}^4$. 
To relax the $\ell_2$ fitting cost to the noise power, we set the noise relaxing term as: 
\begin{align}
 \epsilon_e=\sum\nolimits_{i=1}^{I}P\sigma_{i}^2-2\sqrt{\sum\nolimits_{i=1}^{I}P\sigma_{i}^4}.
\end{align} 
Here, the standard deviation is subtracted twice, because: i) this makes the probability, that the $\ell_2$ fitting cost to be larger than $\parallel \mathbf{e}\parallel^2$, to be very small;  
when the $\ell_2$ fitting cost is allowed to be larger than $\parallel \mathbf{e}\parallel^2$, the minimization of $| \mathbf{s} |$ will distort the source signal; here we favor less source signal distortion at the price of less noise reduction, and
ii) the minimization of $| \mathbf{s} |$ tends to make the residual noise in the estimated source signals sparse. The sparse noise is perceptually notable even if the noise power is low.    
As a result, some perceptible noise remains in the estimated source signal. This method needs only an estimation of the single-channel noise auto-PSD, but not the cross-PSD among microphones or among frames.  Note that a directional noise cannot be considered as a source, since the method depends on the spectral sparsity of the source signal. 

Besides, the $\ell_2$ fit should also be relaxed with respect to the CTF approximation error and the CTF filter perturbations. 
The tolerance is akin to the energy of the noise-free signal, which can be estimated by spectral subtraction as: 
\begin{align}\label{eq:ss}
\hat{\Gamma}_s=\max(\parallel \mathbf{x}\parallel^2-\sum\nolimits_{i=1}^{I}P\sigma_{i}^2, \ 0).
\end{align} 
Empirically, the tolerance with respect to the noise-free signal is set to $\epsilon_s=0.01\hat{\Gamma}_s$. Overall, the tolerance is set to $\epsilon= \epsilon_e+\epsilon_s$.

Thanks to the sparsity constraint, the optimization problem (\ref{eq:classo}) is feasible for (over-)determined configurations as well as under-determined ones. We refer to this method as CTF-based Constrained Lasso (CTF-C-Lasso). 

\subsection{Convex Optimization Algorithm}\label{ssec:cov}

The optimization algorithm presented in this section mainly follows the principle proposed in \cite{arberet2013}. 
Unlike \cite{arberet2013}, the target optimization problem (\ref{eq:classo}) is carried out in the complex domain, and thus the optimization algorithm is also complex-valued.  
The optimization problem consists of an $\ell_1$-norm minimization and  a quadratic constraint, which are both convex.  
The difficulty of this convex optimization problem is that the $\ell_1$-norm objective function is not differentiable. 

The constrained optimization problem (\ref{eq:classo}) can be recast as the following unconstrained optimization problem
\begin{align}\label{eq:pro}
\mathop{\textrm{min}}_{\mathbf{s}}   | \mathbf{s} |+ \iota_C(\mathbf{s}),  
\end{align}
where $C$ denotes the convex set of signals verifying the constraint, $C=\{\mathbf{s} \ |\parallel \mathcal{A}\star\mathbf{s} - \mathbf{x} \parallel^2 \le \epsilon \}$, and $\iota_C(\mathbf{s})$ denotes the indicator function of $C$, 
namely $\iota_C(\mathbf{s})$ equals 0 if $\mathbf{s}\in C$, and $+\infty$ otherwise. 
This unconstrained problem consists of two lower semi-continuous, non-differentiable (non-smooth), convex functions. For this problem, the \emph{Douglas-Rachford} splitting method \cite{combettes2007} is suitable,  
which is an iterative method.  At each iteration, the two functions are split, and their proximity operators $\text{Prox}_{\iota_C(\cdot)}$ and $\text{Prox}_{\gamma| \cdot |}$ (see below) are individually applied. 
The \emph{Douglas-Rachford} method does not require the differentiability of any of the two functions, and is a generalization of the \emph{proximal splitting} method \cite{combettes2005}. Algorithm \ref{alg:dr} summarizes the \emph{Douglas-Rachford} method. 
Here $\alpha$ and $\gamma$ are set as constant values over iterations, e.g.  1 and 0.01 respectively in our experiments. The initialization of $\mathbf{s}_0$ is set as the matrix composed of $J$ replication of the first microphone signal. 
The convergence criteria is set to check if the optimization objective is almost invariant from one iteration to the next. The threshold $\eta_{1}$ is set to $0.01$ in our experiments. In addition, the maximum number of iterations is set to 20.

The proximity operator plays  the most important role in the optimization of  nonsmooth functions. In Hilbert space, the proximity of a complex-valued function $f$ is
\begin{align}
 \text{Prox}_{f}(\mathbf{z}) = \mathop{\textrm{argmin}}_{\mathbf{y}} f(\mathbf{y}) + \parallel \mathbf{z}-\mathbf{y}\parallel^2.
\end{align}
The proximity operator of the $\ell_1$-norm $\gamma| \cdot |$ at point $\mathbf{z}$, aka the shrinkage operator, is given entry-wise by
\begin{align}
 y_i=\frac{z_i}{|z_i|}\text{max}(0,|z_i|-\gamma).
\end{align}

\begin{algorithm}[t] \caption{\label{alg:dr} Douglas-Rachford}
\begin{algorithmic} 
 \STATE Initialization: $l=0$, $\mathbf{s}_0\in\mathbb{C}^{I\times P}, \alpha\in(0,2), \gamma>0 $,
 \REPEAT 
 \STATE  $\mathbf{z}_l = \text{Prox}_{\iota_C(\cdot)}(\mathbf{s}_l)$
 \STATE  $\mathbf{s}_{l+1}=\mathbf{s}_l+\alpha(\text{Prox}_{\gamma | \cdot |}(2\mathbf{z}_l-\mathbf{s}_l)-\mathbf{z}_l)$
 \STATE  $l=l+1$
 \UNTIL $||\mathbf{s}_{l}|-|\mathbf{s}_{l-1}||/|\mathbf{s}_{l}|<\eta_{1}$ 
  \end{algorithmic}
\end{algorithm}

\begin{algorithm}[t] \caption{\label{alg:prox-indic} $\text{Prox}_{\iota_C(\cdot)}(\mathbf{s})$}
\begin{algorithmic} 
 \STATE Input: $\mathbf{x}$, $\mathcal{A}$, $\mathcal{A}^*$, $\mathbf{s}$  
 \STATE Initialization: $l=0$, $\mathbf{u}_{0}=\mathbf{x}$, $\mathbf{p}_0=\mathbf{s}$, $t_0=1$, $\mu\in(0,2/\nu)$ 
 \REPEAT 
 \STATE 1. $l=l+1$
 \STATE 2. $\mathbf{u}_{l}=\mu(\mathbf{I}-\text{Prox}_{\iota_{\parallel \cdot \parallel^2 \le \epsilon}})(\mu^{-1}\mathbf{u}_{l-1}+\mathcal{A}\star\mathbf{p}_{l-1}-\mathbf{x})$
 \STATE 3. $t_l=(1+\sqrt{(1+4t_{l-1}^2)})/2$ 
 \STATE 4. $\tilde{\mathbf{u}}_{l}=\mathbf{u}_{l-1}+\frac{t_{l-1}-1}{t_l}(\mathbf{u}_{l}-\mathbf{u}_{l-1})$
 \STATE 5. $\mathbf{p}_l=\mathbf{s}-\mathcal{A}^*\star \tilde{\mathbf{u}}_{l}$ 
 \UNTIL $\parallel \mathcal{A}\star\mathbf{p}_k - \mathbf{x} \parallel^2 \le 1.1\epsilon$ 
 \STATE Output: $\mathbf{p}_l$
  \end{algorithmic}
\end{algorithm}

The proximity of the indicator function $\iota_C(\mathbf{s})$ is the \emph{projection} of $\mathbf{s}$ onto $C$. 
To compute this proximity, based on the \emph{proximal splitting} method and the Fenchel-Rockafellar duality \cite{rockafellar2015}, an iterative method was derived in \cite{fadili2009}, and used in \cite{arberet2013}. 
However, this method converges linearly, which is very slow especially when the convex set $C$ (also $\epsilon$) is small. As hinted in \cite{fadili2009}, it can be accelerated to the squared speed via the Nesterov's scheme \cite{nesterov2007,beck2009}. 
The accelerated method is summarized in Algorithm \ref{alg:prox-indic}. The acceleration procedure is composed of Step 3 and 4, which are based on the derivation in \cite{beck2009}. 
Here $\mathcal{A}^*$ is the adjoint matrix of $\mathcal{A}$, and is obtained by conjugate transposing the source and channel indices, and then temporally reversing the filters. 
Here $\nu$ is the tightest frame bound of the quadratic operation in the indicator function, and thus is the largest spectral value of the frame operator $\mathcal{A}^* \circ\mathcal{A}$. The power iteration method is used to compute $\nu$, which is summarized in Algorithm~\ref{alg:power-iteration}. 
We set $\mu$ as a constant value over iterations, e.g. $1/\nu$ in the experiments. In Step 2, the \emph{projection} of a variable  $\mathbf{u}$ onto the convex set $\{\mathbf{v} \ |\parallel \mathbf{v} \parallel^2 \le \epsilon \}$ can be easily obtained as 
\begin{align}
 \text{Prox}_{_{\iota_{\parallel \cdot \parallel^2 \le \epsilon}}}(\mathbf{u}) = \text{min}(1,\frac{\sqrt{\epsilon}}{\parallel \mathbf{u} \parallel})\mathbf{u}.
\end{align}
In Algorithm \ref{alg:prox-indic}, the variable $\mathbf{p}_k$ iteratively moves from the initial point $\mathbf{s}$ to its \emph{projection}, 
thence a convergence criteria is set to check the feasibility of the constraint. The slack factor $1.1$ is set to avoid the time consuming long tail of convergence, which  however leads to a possible small bias of the $\ell_2$-norm constraint. In addition, the maximum number of iterations is set to 300.

\begin{algorithm}[t] \caption{\label{alg:power-iteration} Power Iteration}
\begin{algorithmic} 
 \STATE Input: $\mathcal{A}$, $\mathcal{A}^*$ 
 \STATE Initialization: $\mathbf{v}\in\mathbb{C}^{J\times P}$ 
 \REPEAT 
 \STATE $\mathbf{w}={\mathcal{A}^*}\star(\mathcal{A}\star\mathbf{v})$ 
 \STATE $\mathbf{v}=\mathbf{w}/\parallel \mathbf{w} \parallel$ 
 \UNTIL convergence
 \STATE Output: $\nu=\parallel \mathbf{w} \parallel$
  \end{algorithmic}
\end{algorithm}

\section{Experiments}\label{sec:experiments}

In this section, we evaluate the quality of the estimated source signals, in terms of the performance of source separation, speech dereverberation and noise reduction.

\subsection{Experimental Configuration}

\subsubsection{Dataset}
The multichannel impulse response data \cite{hadad2014} is used, which  
was recorded using a 8-channel linear microphone array in the speech and acoustic lab of Bar-Ilan University, with room size of $6$~m $\times$ $6$~m $\times$ $2.4$~m. 
The reverberation time is controlled by 60 panels covering the room facets. In the reported experiments, we used the recordings with $T_{60}=0.61$~s. The RIRs are truncated to correspond to $T_{30}$, and have a length of 5600 samples. 
The speech signals from the TIMIT dataset \cite{garofolo1988} are taken as the source signals, with a duration of about 3~s. 
TIMIT speech is convolved with a RIR as the image of one source. 
Multiple  image sources are summed up.  
For one such mixture, the source direction and the microphone-to-source distance of each source are randomly selected from $-90^{\circ}$:$15^{\circ}$:$90^{\circ}$ and \{1 m, 2 m\}, respectively.
Note that the mutiple sources consist of different TIMIT speech utterances and different impulse responses in terms of source directions. 
To generate noisy microphone signals, a spatially uncorrelated stationary speech-like noise is added to the noise-free mixture, the noise level is controlled by a wide-band input signal-to-noise ratio (SNR). Note that SNR refers to the averaged single source-to-noise ratio over multiple sources. 
To evaluate the robustness of the methods to the perturbations of the RIRs/CTFs, 
a proportional random Gaussian noise is added to the original filters $a^{i,j}(n)$ in the time domain to generate the perturbed filters denoted as $\tilde{a}^{i,j}(n)$.
The perturbation level is denoted as the normalized projection misalignment (NPM) \cite{morgan1998} in decibels (dB). 
Various acoustic conditions in terms of the number of microphones and sources, SNRs, and NPMs are tested. For each condition, 20 runs are executed, and the averaged performance measures  are computed.

\subsubsection{Performance Metrics}
The signal-to-distortion ratio (SDR) \cite{vincent2006} in dB is used to evaluate the overall quality of the outputs. The unprocessed microphone signals are evaluated as the baseline scores. The overall outputs, i.e. (\ref{eq:out}) for CTF-MINT and CTF-MPDR, and (\ref{eq:classo}) for CTF-C-Lasso, are evaluated as the output scores. 

The signal-to-interference ratio (SIR) \cite{vincent2006} in dB is specially used to  evaluate the source separation performance. This metric focuses on the suppression of interfering sources, thence the additive noise would be eliminated. The unprocessed noise-free mixtures, i.e. $\sum_{j=1}^J a_{p}^{i,j} \star s_{p}^j$, are evaluated as the baseline scores.
For CTF-MINT and CTF-MPDR, we can simply take the noise-free output, i.e. $\sum_{i=1}^{I} h_p^{i} \star (\sum_{j=1}^J a_{p}^{i,j} \star s_{p}^j)$ in (\ref{eq:out}), for evaluation. However, for CTF-C-Lasso, we have to test the overall outputs, since the noise-free output is not available. Experimental results show that CTF-C-Lasso has low residual noise, thus the SIR measure is assumed  not to be significantly influenced by the output additive noise. 

The perceptual evaluation of speech quality (PESQ) \cite{rix2001} is specially used to  evaluate the dereverberation performance. The interfering sources and noise would be eliminated. For each source, its unprocessed image sources, i.e. $a_{p}^{i,j} \star s_{p}^j$ are evaluated as the baseline scores.    
For CTF-MINT and CTF-MPDR, the noise-free single source output, i.e. $\sum_{i=1}^{I} h_p^{i} \star (a_{p}^{i,j} \star s_{p}^j)$ is evaluated. For CTF-C-Lasso, again we have to test the overall outputs. However, the residual interfering sources and noise affect the PESQ measure to a large extent.
Therefore, we should note that the PESQ scores of CTF-C-Lasso are highly underestimated. 

The output SNR in dB is used to evaluate the noise reduction performance. 
The input SNR is taken as the baseline scores. 
For CTF-MINT and CTF-MPDR, the output SNR is computed as the power ratio between the noise-free outputs and the output noise, i.e. $\sum_{i=1}^{I} h_p^{i} \star e_p^i$. For CTF-C-Lasso, the noise PSDs in the output signals are first blindly estimated using the method proposed in \cite{li2016icassp}. The power of the noise-free outputs are estimated by spectral subtraction following the principle in (\ref{eq:ss}), and then the output SNR is obtained by taking the ratio of them. It is shown in \cite{li2016icassp} that the estimation error of noise PSD is around 1 dB, thence the estimated output SNRs are reliable. 

SDR, SIR and PESQ are evaluated in the time domain, thence the signals mentioned above are actually their corresponding time-domain signals reconstructed using inverse STFT.  
The output SNR for CTF-MINT and CTF-MPDR are computed either in the time domain or in the STFT-domain, while the output SNR for CTF-C-Lasso is computed in the STFT domain. 

\begin{figure*}[t]
\centering
{\includegraphics[height=0.23\textwidth]{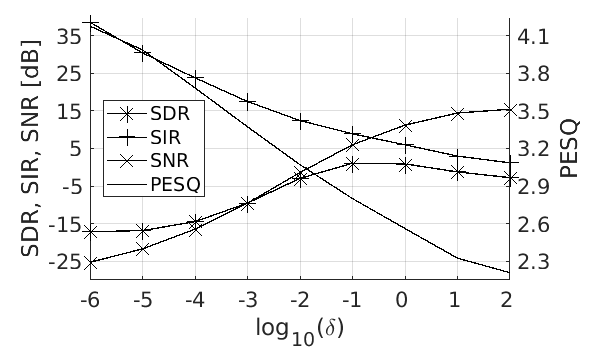}} \hspace{0.7cm}
{\includegraphics[height=0.23\textwidth]{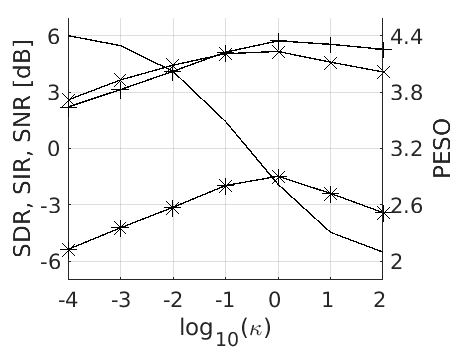}} 
\caption{The performance measures as a function of  $\delta$ for CTF-MINT (\emph{left}) and  $\kappa$ for CTF-MPDR (\emph{right}). $I=4$ and $J=3$. The input SNR is 10 dB. SDR, SIR and PESQ of the unprocessed signals are -6.9 dB, -3.0 dB and 1.85, respectively. Two vertical axes are used due to the different scales and units of the performance measures.} 
\label{fig:regu}
\vspace{-0.0cm}
\end{figure*}

\subsubsection{Parameter Settings}
The sampling rate is 16 kHz. The STFT is calculated using a Hamming window, with window length and frame step of $N=1,024$ (64~ms) and $D=N/4=256$, respectively. 
The CTFs are computed from the time-domain filters using (\ref{eq:apk}). 
The CTF length $L_a$ is 29.
For the over-determined recording system, i.e. $I>J$, the length of the inverse filter of CTF-MINT, i.e. $L_h^{\text{mint}}$, is computed via (\ref{eq:lhmint}) with $\rho=1$, which makes $\mathbf{A}$ square. 
Pilot experiments show that a longer inverse filter (or a larger $\rho$) does not noticeably improve the performance measures, while leading to a larger computational cost. 
For the case of $I\le J$,  $\rho$ is set to be less than and close to $\frac{I}{J}$, and $\rho$ should be small to avoid an unreasonable long inverse filter. 
The exact values of $\rho$ will be given in the following experiments depending on the specific values of $I$ and $J$. 
The length of the inverse filter of CTF-MPDR, i.e. $L_h^{\text{mpdr}}$, 
is computed via (\ref{eq:lhmpdr}) with $\varrho=1$, thus $\mathbf{A}^{j_d}$ is square.  The optimal setting of the modeling delay in $\mathbf{d}$ is related to the length of the inverse filters. In the experiments, it is respectively set to 6 and 3 taps for CTF-MINT and CTF-MPDR as a good tradeoff for the different inverse filter lengths in various acoustic conditions. 

Thanks to the normalization factors in (\ref{eq:ctf-mint}) and (\ref{eq:ctf-mpdr}), the same regularization factors $\delta$ and $\kappa$ are suitable for all frequencies. Moreover, they are robust to any possible numerical scales of the filters and the signals in different datasets.   
Fig.~\ref{fig:regu} shows the performance measures of CTF-MINT and CTF-MPDR as a function of $\delta$ and $\kappa$, respectively. For CTF-MINT,  
with the increase of $\delta$, the inaccuracy of inverse filtering increases, while the energy of the inverse filters decreases. 
From the \emph{left} plot of Fig.~\ref{fig:regu}, it is observed that the output SNR gets larger with the increase of $\delta$, which confirms that the additive noise can be suppressed by decreasing the energy of the inverse filter. 
However, SIR and PESQ scores become smaller with the increase of $\delta$ due to the larger inaccuracy of inverse filtering, which leads to more residual interfering sources and reverberation. 
Integrating these effects, SDR first increases then decreases with the increase of $\delta$. 
In a similar way, the energy of the inverse filters also affects the robustness of the inverse filtering to the CTF perturbations. 
In summary, we consider two representative choices of $\delta$:  i) a relatively small one, i.e. $10^{-5}$, leads to an accurate inverse filtering but a large energy of the inverse filter; 
this is suitable for the case where both the microphone noise and the CTF perturbations are small, and ii) a large one, i.e. $10^{-1}$, achieves an output SNR being slightly larger than the input SNR thus avoiding the amplification of the additive noise. In the following experiments, the former is used for the noise-free case, and the latter is used for the noisy case.  
This partially oracle configuration is a bit unrealistic, but is useful to show the full potential of CTF-MINT. See \cite{kodrasi2013} for further discussion on the optimal setting of $\delta$.

For CTF-MPDR, $\kappa$ controls the tradeoff between the distortionless of the desired source and the power of the output. 
The minimization of the power of the output will suppress both the  interfering sources and the noise. 
From the \emph{right} plot of Fig.~\ref{fig:regu}, we observe that PESQ decreases along with the increase of $\kappa$, due to the increased distortions of the desired source. 
SIR and output SNR can be increased by increasing $\kappa$ until $\kappa=1$. A larger $\kappa$, e.g. $10^2$, leads to a smaller SIR and output SNR although the power of the output is smaller, since the desired signal is also heavily distorted and suppressed. Overall, $\kappa$ is set to $10^{-1}$, which achieves a high PESQ score and good other measures.

\begin{figure*}[t]
\centering
{\includegraphics[width=0.32\textwidth]{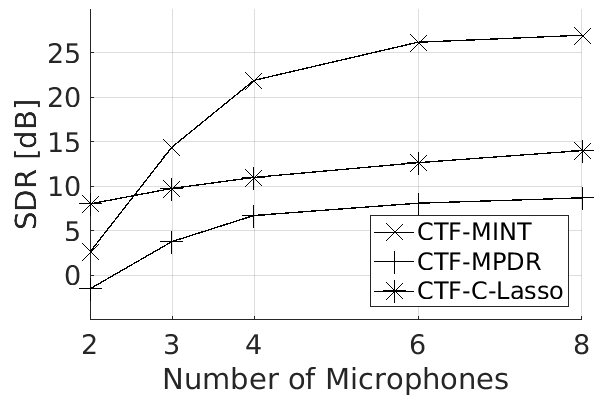}}
{\includegraphics[width=0.32\textwidth]{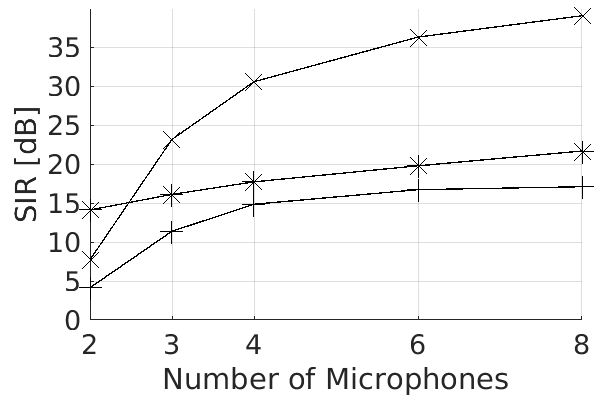}} 
{\includegraphics[width=0.32\textwidth]{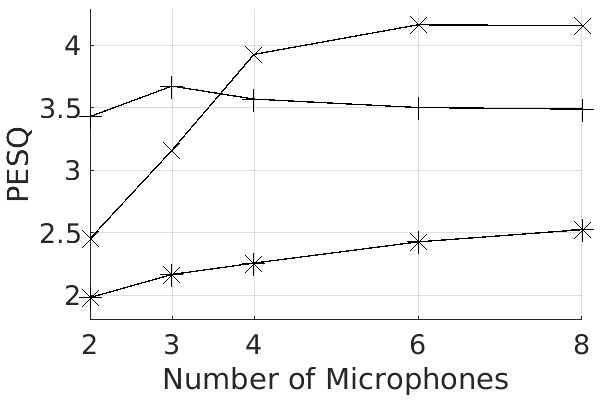}} 
\caption{The performance measures as a function of the number of microphones, $J=3$. The microphone signals are noise free. SDR, SIR and PESQ of the unprocessed signals are -6.9 dB, -3.0 dB and 1.85, respectively. Note that the legends in this figure are common to all the following figures. } 
\label{fig:mics}
\vspace{-0.0cm}
\end{figure*}

\begin{figure*}[t]
\centering
{\includegraphics[width=0.32\textwidth]{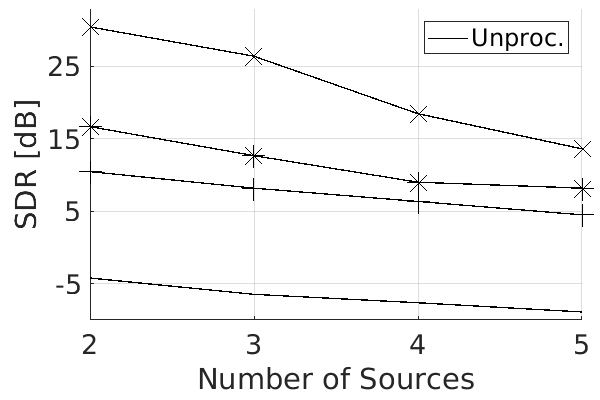}}
{\includegraphics[width=0.32\textwidth]{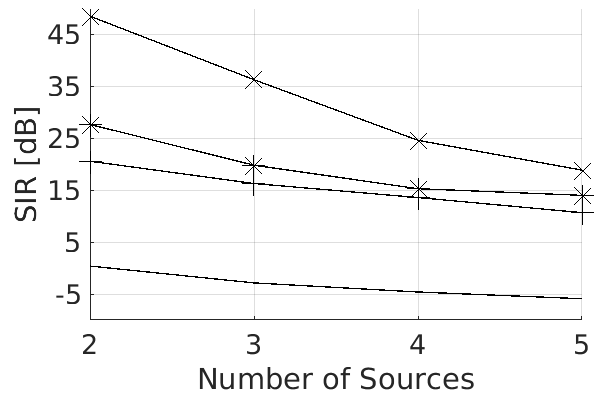}} 
{\includegraphics[width=0.32\textwidth]{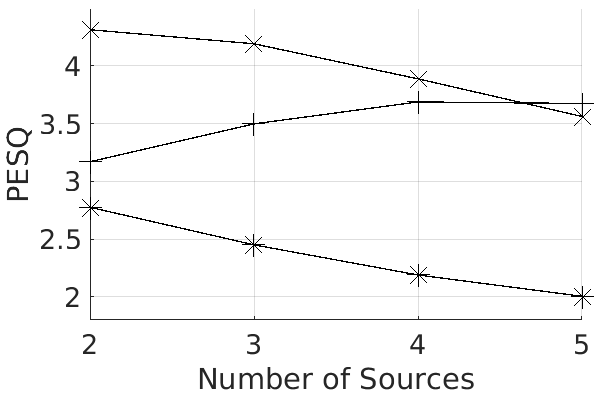}} 
\caption{The performance measures as a function of the number of sources, $I=6$. The microphone signals are noise free. PESQ of the unprocessed signals is 1.85.} 
\label{fig:sou}
\vspace{-0.0cm}
\end{figure*}

\begin{figure*}[t]
\centering
{\includegraphics[width=0.245\textwidth]{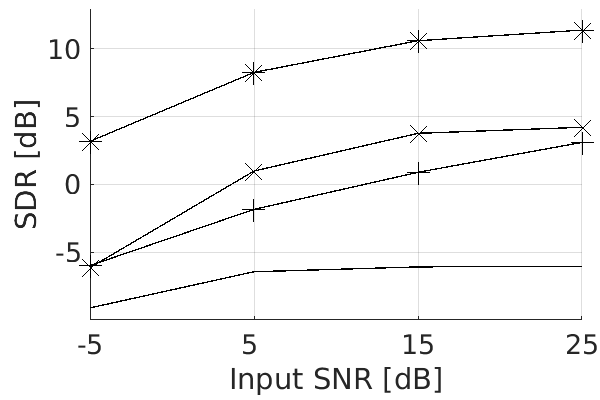}}
{\includegraphics[width=0.245\textwidth]{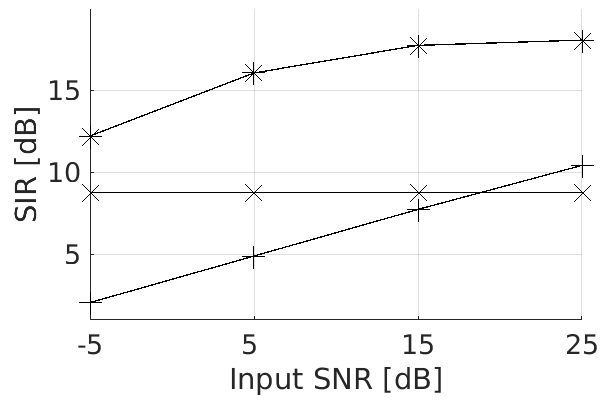}} 
{\includegraphics[width=0.245\textwidth]{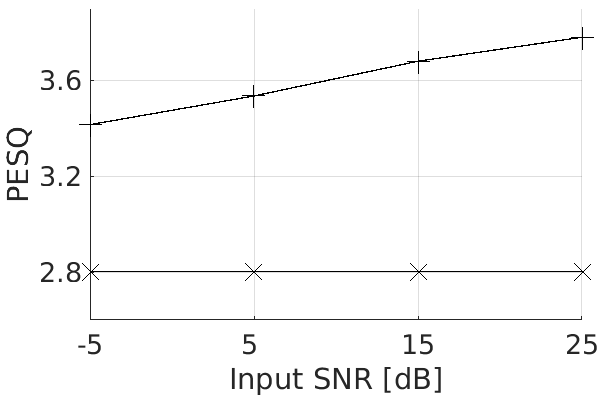}} 
{\includegraphics[width=0.245\textwidth]{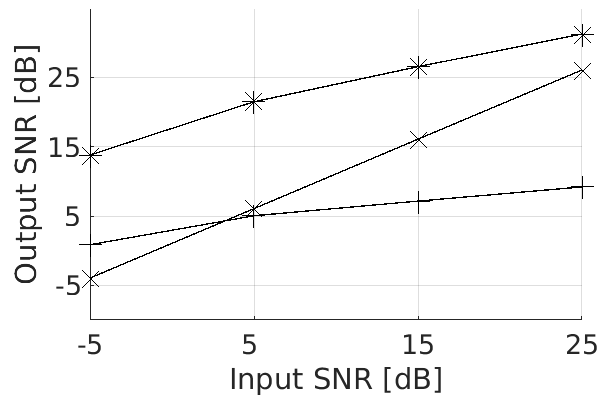}} 
\caption{The performance measures as a function of input SNRs, $I=4$ and $J=3$. SIR and PESQ of the unprocessed signals are -3.0 dB and 1.85, respectively. PESQ for CTF-C-LASSO is not shown since it is inaccurate due to the residual noise.} 
\label{fig:snr}
\vspace{-0.0cm}
\end{figure*}

\subsection{Influence of the Number of Microphones}
Fig.~\ref{fig:mics} shows the results as a function of the number of microphones.  The source number is fixed to three. 
In this experiment, the microphone signals are noise free, thus the output SNR is not reported. 
For CTF-MINT, $\rho$ is set to 0.55 and 0.8 for the cases of two and three microphones, respectively. Consequently the length of the inverse filters are about five times the CTF length.  

For CTF-MINT, the scores of all the three metrics dramatically decrease when the number of microphones goes from four to three and to two, namely from the over-determined case to the determined case and to the under-determined case. 
This indicates that the inaccuracy of the inverse filtering is large for the non over-determined case, due to the insufficient degrees of freedom of the inverse filters as spatial parameters. 
CTF-MPDR suppresses the interfering sources by minimizing the power of the output, and implicitly also by the inverse filtering with a target of zero signal. Therefore, as for CTF-MPDR, the metrics to measure the interfering sources suppression performance, i.e. SDR and SIR, also significantly degrade for the non over-determined case. Along with the increase of number of microphones,   the PESQ score slightly varies, which means that the inverse filtering of the desired source is not considerably affected, due to the small variation of the output power.  The performance measures of CTF-C-Lasso increases almost linearly  with the growing number of microphones, no matter whether it is under-determined or over-determined, thanks  to exploiting the spectral sparsity.    
For the over-determined case, i.e. four microphones or more, SDR and SIR for the three methods slowly increase with the growing number of microphones, and CTF-MINT has a larger changing rate. CTF-C-Lasso achieves the worst PESQ score due to the influence of the residual interfering sources. By listening to the outputs of CTF-C-Lasso, they are not perceived as more reverberant.

Overall, without considering the noise reduction, CTF-MINT performs the best for the over-determined case. For instance,  CTF-MINT achieves an SDR of 21.9 dB by using four microphones, which is a very good source recovery SDR score. 
CTF-C-Lasso performs the best for the under-determined case. For instance,  CTF-C-Lasso achieves an SDR of 8.4 dB by using only two microphones.
By  only using the mixing filters of one source, the source separation performance of CTF-MPDR is worse than the other two methods.

\subsection{Performance for Various Number of Sources}

Fig.~\ref{fig:sou} shows the results as a function of the number of sources. In this experiment, the number of microphones is fixed to six.
The microphone signals are noise free, thus the output SNR is not reported. 
From this figure, we can observe that the performance measures of the three methods degrade with the increase of the number of sources, except for the PESQ score of CTF-MPDR. CTF-MINT achieves the best performance, even if it exhibits the largest performance degradation. This is somehow consistent with the experiments with various number of microphones that good performance requires a large ratio between the number of microphones and the number of sources. 
Both CTF-MPDR and CTF-C-Lasso  have  smaller performance degradation. 
At first sight, it is surprising that CTF-MPDR achieves a larger PESQ score when more sources are present in the mixture. The reason is that the normalized output power, i.e.  $\frac{\phi_a^{j_d}}{\phi_x} \parallel \mathbf{X}\mathbf{h} \parallel^2$, becomes smaller with the increase of the number of sources due to a larger ${\phi_x}$. Correspondingly, the inverse filtering inaccuracy of the desired source, i.e. $\parallel \mathbf{A}^{j_d}\mathbf{h}-\mathbf{d} \parallel^2$, becomes smaller as well. 

\begin{figure*}[t]
\centering
{\includegraphics[width=0.32\textwidth]{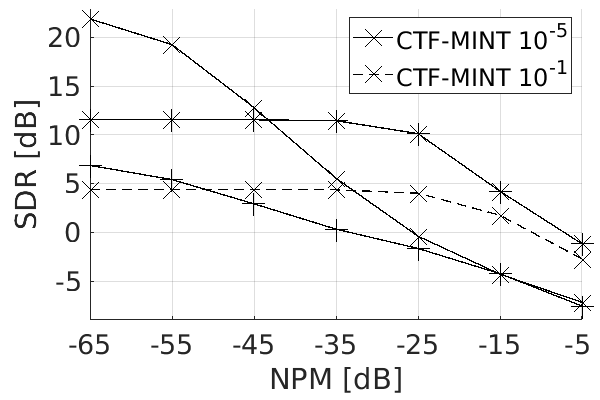}}
{\includegraphics[width=0.32\textwidth]{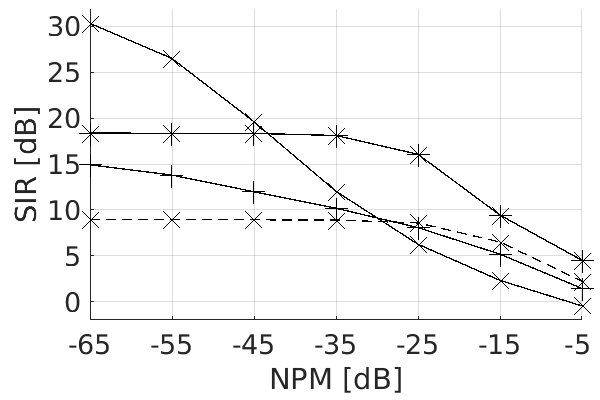}} 
{\includegraphics[width=0.32\textwidth]{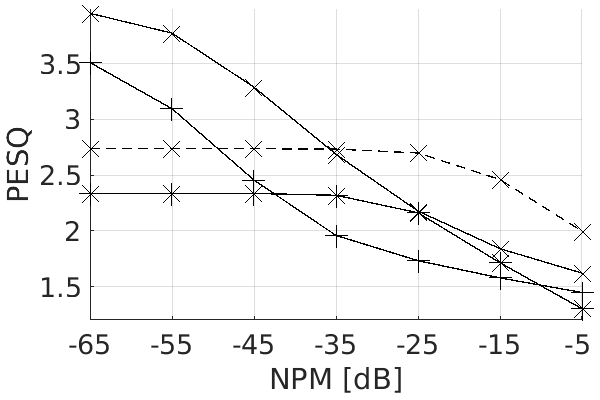}} 
\caption{The performance measures as a function of NPM, $I=4$ and $J=3$. The microphone signals are noise free. SDR, SIR and PESQ of the unprocessed signals are -6.9 dB, -3.0 dB and 1.85, respectively.} 
\label{fig:npm}
\vspace{-0.0cm}
\end{figure*}

\setlength{\tabcolsep}{3.0pt}
\begin{table*}[t]
\centering
\caption{ The SDR scores and the computation times  for six representative acoustic conditions. The SDR scores of the unprocessed signals are given in the previous experiments. }
\label{tab:cdr-ct}
\begin{tabular}{c c c c | c c  c c  c c| c c  c c c c }  
\multicolumn{4}{c|}{Acoustic Condition} & \multicolumn{6}{c|}{SDR [dB]} & \multicolumn{6}{c}{Computation Time per Mixture [s]} \\  
{\scriptsize $I$} & {\scriptsize $J$} & {\scriptsize SNR}  & {\scriptsize NPM}  & {\scriptsize CTF-MINT} & {\scriptsize CTF-MPDR } & {\scriptsize CTF-C-Lasso} & {\scriptsize LCMP} & {\scriptsize TD-MINT}  & {\scriptsize W-Lasso} & {\scriptsize CTF-MINT} & {\scriptsize CTF-MPDR} & {\scriptsize CTF-C-Lasso} & {\scriptsize LCMP} & {\scriptsize TD-MINT} & {\scriptsize W-Lasso} \\  \hline 
4   &  3  &-    & -   &  21.9    &   6.7    & 11.0    & -3.6 & - & 18.9    &  25.4    &   4.9    & 1987       & 1.1 & - & 4284               \\
6   & 2 & - & - & 30.4 & 10.4 & 16.6 & -0.3 & 30.0 & 31.2  & 5.8 & 4.2  & 1688 & 1.1 & 142 & 3843  \\
6   & 3   &-    & -   &26.3      & 8.2      & 12.6        &-0.6 & - & 23.8    &12.2      & 5.9      & 2827        &1.2 & - & 5961             \\
6   & 5   & -   & -   &13.6      &4.5       &8.2          &-6.4 & -  & 14.7        &229.6      &12.4       &5679          &1.9 & -  &  10134         \\
4   &  3  & 15 dB  &-    &3.8       &0.9      &10.6         &-14.7 &- & -        &21.9       &6.7      &1500         &1.1 &  - &    -         \\
4   &  3  &  -  & -15 dB &1.7       &-4.3      &4.2          &-4.1 & -  & 0.5        &21.9       &6.7     &1440         &1.1 & - &  4245              \\
\end{tabular}
\vspace{-.0cm}
\end{table*}

\subsection{Influence of Additive Noise}

Fig.~\ref{fig:snr} shows the results as a function of the input SNR. The number of microphones and of sources are respectively fixed to four and three.
As mentioned above, for the noisy case, the regularization factor $\delta$ is set to $10^{-1}$. The inverse filter of CTF-MINT is invariant for various input SNRs, since it depends only on the CTF filters, but not on the microphone signals. As a result, the SIR and PESQ scores are constant, but are much smaller than the noise-free case with $\delta=10^{-5}$, see Fig.~\ref{fig:mics}.
The SNR improvement is also a constant value, about 1 dB. 
For CTF-MPDR,  SIR and PESQ  are smaller when the input SNR is lower, since a larger input noise leads to a larger output noise, thus degrades the suppression of the interfering sources, and distorts the inverse filtering of the desired source. Along with the increase of the input SNR, the output SNR increases, but the SNR improvement decreases.  The SNR improvement is negative when the input SNR is larger than 5 dB, which means the microphone noise is amplified. 
For CTF-MINT and CTF-MPDR, the residual noise is significant, which indicates that the inverse filtering is not able to efficiently suppress the white noise.
Therefore, a single channel noise reduction process is needed as a postprocessing, as in \cite{cohen2003integrated,gannot2004speech}.
The output SNR of CTF-C-Lasso is always larger than the input SNRs, which means that the microphone noise is efficiently reduced. SDR and SIR of CTF-C-Lasso degrades for the low SNR case, but not much.

\subsection{Influence of CTF Perturbations}

Fig.~\ref{fig:npm} shows the results as a function of NPMs. For CTF-MINT, two choices of the regularization factor, i.e. $10^{-5}$ and $10^{-1}$, are tested. 
As expected, all the metrics become worse with the increase of NPM, thus we only analyze the SDR scores. Note that, when NPM is -65 dB, the three methods achieve almost the same performance measures as with the perturbation-free case.
Along with the increase of NPMs, the performance of CTF-MINT with $\delta=10^{-5}$ dramatically degrades from a large score to a very small score, which indicates its high sensitivity to CTF perturbations. 
In contrast, CTF-MINT with $\delta=10^{-1}$ has a small performance degradation rate, but the performance is poor even for the low NPM case. 
The performance measures of CTF-MPDR almost linearly decreases with a relatively large degradation rate. 
The performance of CTF-C-Lasso is stable until NPM equals -35 dB, and quickly degrades when NPM is larger than -25 dB. 

In CTF-MINT, the inverse filter is designed to respectively satisfy the targets of desired source and interfering sources. 
Therefore, the CTF perturbations of the desired source will not significantly affect the suppression of interfering sources, and vice versa. 
Moreover, in CTF-MPDR, the inverse filter is computed  depending only on the CTFs of the desired source, thence the CTF perturbations of the interfering sources will not affect the inverse filtering at all.
In contrast, in CTF-C-Lasso, all sources are simultaneously recovered based on the CTFs of all of them, consequently the CTF perturbations of one source will affect the recovery of all sources. These assertions have been verified by some pilot experiments. 

\subsection{Comparison with Baseline Methods}

To benchmark the proposed methods, we compare them with three baseline methods:
\begin{itemize}
\item LCMP beamformer \cite{van2004} based on the narrowband assumption. 
Based on the steering vectors and the correlation matrix of microphone signals, a beamformer is computed to preserve one desired source and zero out the others, and to minimize the power of the output. The RIRs are longer than the STFT window, thus the steering vector should be computed as the Fourier transform of the truncated RIRs. In this experiment, the steering vector is set to the CTF tap with the largest power.
\item Time domain MINT (TD-MINT) \cite{hikichi2007}. This method is also set to recover the direct-path source signal with an energy regularization.  In this experiment, we extend this method to the multisource case. We only test the condition with $I=6$ and $J=2$, following the principle of the proposed method, the length of inverse filter and the modeling delay are set to 2800 and 1024,  respectively. Other conditions require too long inverse filters that cannot be implemented within basic memory ressources on a personal computer.
\item Wideband Lasso (W-Lasso) \cite{kowalski2010}. The regularization factor is set to $10^{-5}$, which is empirically suitable for the noise-free case.
\end{itemize}

Table~\ref{tab:cdr-ct} presents the SDR scores for six representative acoustic conditions, as well as the computation times which  will be analyzed in the next section. Note that `-'  means noise-free and perturbation-free in the columns of SNR and NPM, respectively. LCMP performs poorly for all conditions, which verifies the assertion that the narrowband assumption is not suitable for the long RIR case. 
CTF-MINT achieves a bit higher SDR score than TD-MINT, despite the fact that the CTF-based filtering is an approximation of the time-domain filtering. This is mainly due to much shorter filters in the STFT/CTF domain.
W-Lasso noticeably outperforms CTF-C-Lasso for the noise-free and perturbation-free cases, due to its exact time-domain convolution. 
W-Lasso has a similar noise reduction capability with CTF-C-Lasso, however the regularization factor is difficult to set for a proper noise reduction, thence the results of W-Lasso for the noisy case is not reported.  
Compared to CTF-C-Lasso, W-Lasso has a larger performance degradation rate with the increase of the number of sources and of filter perturbations.   

\subsection{Analysis of Computational Complexity}

Table~\ref{tab:cdr-ct} also presents the averaged computation time for one mixture with a duration of 3~s. All methods were implemented in MATLAB.
CTF-MINT and CTF-MPDR  computation times comprise the inverse filters computation and the inverse filtering on the microphone signals, and the former dominates the computation time.
From (\ref{eq:sctf-mint}) and (\ref{eq:sctf-mpdr}), the computations include the multiplication and inversion of the matrices, thence the complexity is cubic in matrix dimension.
We consider square matrices $\mathbf{A}$ in (\ref{eq:sctf-mint}) and $\mathbf{A}^{j_d}$ in (\ref{eq:sctf-mpdr}), whose dimension is equal to $IL_h$. 
From (\ref{eq:lhmint}) and (\ref{eq:lhmpdr}), $IL_h$ is proportional to the filter length $L_a$,  to $\frac{I-J}{IJ}$ for CTF-MINT, and to $\frac{I}{I-1}$ for CTF-MPDR. The inverse filters are respectively computed for each source and each frequency. 
Overall, CTF-MINT and CTF-MPDR  have a computational complexity of 
$\mathcal{O}(\frac{KL_a^3I^3J^4}{(I-J)^3})$ and $\mathcal{O}(\frac{KL_a^3I^3J}{(I-1)^3})$, respectively, where $K=N/2+1$ is the number of frequency bins.
 The complexity of TD-MINT can be derived from the complexity of CTF-MINT by replacing the  CTF length with the RIR length and setting $K$ to 1. Since it is proportional to the cube of RIR length, the complexity is prohibitive for most settings. The LCMP beamformer is similar to CTF-MINT, just using an instantaneous steering vector and an instantaneous inverse filter, namely the length of CTF and inverse filter are both 1, thence it has the lowest computation complexity. 
These methods have a close-form solution and thus low computational complexity. These can be  verified by the computation times shown in Table \ref{tab:cdr-ct}.

The iterative optimization of CTF-C-Lasso leads to a high computational complexity. 
Unlike the Newton-style methods employing the second-order derivative, the \emph{Douglas-Rachford} optimization method is a first-order method,  thence the complexity is linear with respect to the problem size, 
specifically the length of microphone signals and filters, and the number of  microphones and sources.  
The most time consuming procedure in Algorithm \ref{alg:dr} is the computation of the  proximity of the indicator function, i.e. the \emph{projection}.
To verify this, we can compare the \emph{Douglas-Rachford} method with the optimization algorithm for the Lasso problem (\ref{eq:lasso}) that does not have an $\ell_2$-norm constraint and thus an indicator function.
In \cite{li2017icassp}, we solved the unconstrained Lasso problem using the fast iterative shrinkage-thresholding algorithm (FISTA) \cite{beck2009}, which is also a \emph{proximal splitting} method just without computing the proximity of the indicator function. 
As reported in \cite{li2017icassp}, FISTA  needs only about tens of seconds per mixture, while here \emph{Douglas-Rachford} needs thousands of seconds per mixture, see Table \ref{tab:cdr-ct}.
As stated in Section \ref{ssec:cov}, in Algorithm \ref{alg:prox-indic}, the variable iteratively moves from the initial point to its \emph{projection} in the $\ell_2$ convex set. 
Therefore, a larger convex set caused by a larger noise power (a larger $\epsilon$) needs less iterations to reach the \emph{projection}, and needs less computation time.  This can be verified by the fact that the case with SNR of 15 dB needs less computation time than the noise-free case. 
When the CTF perturbations is large, e.g. NPM is -15 dB, the optimized objective, i.e. $|\mathbf{s}|$, is large, thence less iterations (and less computation time) are needed to converge. 
The CTF convolution at one frequency has a much smaller data size than the time-domain convolution, as a result,  the CTF-based \emph{Douglas-Rachford} method  only requires of the order of ten iterations to converge, while the time-domain W-Lasso method requires tens of thousands iterations to converge. 
As shown in Table \ref{tab:cdr-ct}, the W-Lasso method needs more computation time than CTF-C-Lasso, although it is unconstrained and optimized by FISTA. 

\section{Conclusion}\label{sec:conclusion}
Three source recovery methods based on CTF have been proposed in this paper. CTF-MINT is an ideal over-determined source recovery method when the microphone noise and mixing filter perturbations are small. 
It has a relative low computational complexity. However, it is sensitive to the microphone noise and filter perturbations. CTF-MPDR is also more suitable for the over-determined case than for the non over-determined case. It achieves the worst performance among the three proposed methods but with the lowest computational cost.
The major virtue of CTF-MPDR is that it only requires the mixing filters of the desired source, which makes it more practical. Thanks to exploiting the spectral sparsity, CTF-C-Lasso is able to perform well in the under-determined case, and to efficiently reduce the microphone noise. 
However, it requires the mixing filters of all sources, which are not easy to obtain in practice. In addition, the computational cost is high due to the iterative optimization procedure.



\end{document}